\documentclass[%
 reprint,
 amsmath,amssymb,
 aps,
]{revtex4-2}

\usepackage{graphicx}% Include figure files
\usepackage{dcolumn}% Align table columns on decimal point
\usepackage{bm}% bold math
\usepackage{xspace}  %for space after degree in tc
\usepackage{circledsteps}
\newcommand{\tc}{\textdegree\xspace}
\begin{document}

%\preprint{APS/123-QED}

%\title{Manuscript Title:\\with Forced Linebreak}% Force line breaks with \\

%\title{Predicting the topography of superhydrophobic surfaces from droplet adhesion}
\title{Probing superhydrophobic surface topography using droplet adhesion}

%Advancing surface characterization: A numerical approach for force-based analysis of pillared superhydrophobic surfaces

\author{Pawan Kumar}
\email{kumar.pawan.cot@gmail.com}
\affiliation{%
 Department of Chemical Engineering, The University of Melbourne, Parkville, Melbourne Victoria 3010, Australia
}%
\author{Marta Krasowska}
\affiliation{Future Industries Institute, The University of South Australia, Mawson Lakes 5095, Australia}
\author{Joseph D. Berry}%
 \email{berryj@unimelb.edu.au}
\affiliation{%
 Department of Chemical Engineering, The University of Melbourne, Parkville, Melbourne Victoria 3010, Australia
}%

\date{\today}% It is always \today, today,

\begin{abstract}
Understanding contact line dynamics on superhydrophobic surfaces with microscopic structures is essential for designing materials with reduced drag, anti-icing, self-cleaning, and anti-fouling properties. Using numerical simulations, we demonstrate that forces on droplets receding over structured surfaces are governed by microscale deformations near the contact line. We present and experimentally validate an expression demonstrating that adhesion force increases logarithmically with pillar area fraction at constant droplet volume and pillar surface chemistry. Furthermore, we establish that the average tensile force measured in direct force measurements provides a more reliable indicator of surface structure than the commonly used maximum force. This newfound insight enables precise quantification of superhydrophobic surface structure using a droplet probe.

\end{abstract}

\maketitle

%\tableofcontents

% \section{\label{sec:level1}First-level heading:\protect\\ The line
% break was forced \lowercase{via} \textbackslash\textbackslash}

Surfaces that can resist wetting by liquids and offer low friction to rolling droplets are critical for low-drag, anti-icing, self-cleaning and anti-fouling applications \cite{cheng_surface_2015,kreder_design_2016,blossey_self-cleaning_2003,rana_surface_2010}. Surfaces with microscopic pillars coated with a hydrophobic chemical layer are known to have remarkable superhydrophobicity, characterized by a high advancing contact angle ($>$150\textdegree) and very low contact angle hysteresis ($<$10\textdegree). The wetting behavior of such surfaces depends on the local density of the pillars, also known as the pillar area fraction ($\phi$) \cite{cassie1944wettability,dorrer2006advancing,yeh2008contact,dubov2014contact,dorrer2009some,humayun2022retention,backholm2020water,wang2025coalescence,harvie_contact-angle_2024}. Parameters such as the advancing and receding contact angles and the adhesion and friction forces between a droplet and the surface depend on $\phi$. It is therefore imperative to have methods that can estimate the local area fraction on such surfaces.

Contact angle goniometry (CAG) is the most commonly used technique for surface characterization due to its simplicity and utility \cite{allred_wettability_2017, huhtamaki_surface-wetting_2018}. By measuring advancing ($\theta_{\rm{a}}$) and receding ($\theta_{\rm{r}}$) contact angles, other droplet parameters such as adhesion and friction forces can be indirectly inferred. However, measurement precision is highly dependent on the precise identification of the baseline and the choice of the fitting algorithm used to trace the contour of the droplet \cite{neumann_techniques_1979, sklodowska_method_1999, law_surface_2016, del_rio_contact_1998, rotenberg_determination_1983, vuckovac_uncertainties_2019, shaw_accurate_2024}. Obtaining a perfect droplet contour becomes extremely challenging for surfaces with extreme wettability, such as highly hydrophobic ($\theta_{\rm{a}}>$150\textdegree \cite{liu_improving_2019}) or highly hydrophilic ($\theta_{\rm{a}}<$40\textdegree \cite{konduru_static_2010}) surfaces. In addition, if surface roughness is comparable to droplet size, identifying a precise baseline is difficult, and contact angle measurements are incorrect \cite{shaw_accurate_2024}. Additionally, CAG only gives qualitative information on the local surface topography. 

Direct force measurement techniques such as drop probe microscopy \cite{hokkanen_forcebased_2021, liimatainen_mapping_2017, pilat_dynamic_2012, gao_how_2018} and atomic force microscopy \cite{daniel_mapping_2019, daniel_quantifying_2020, shi_measuring_2015} have emerged as a promising alternative to CAG. These techniques are well suited for directly measuring friction and adhesion forces on surfaces with complex topography and have been used for detailed qualitative surface characterization \cite{liimatainen_mapping_2017, daniel_mapping_2019}. 
%Although force-based methods provide valuable insight into surface-wetting properties and are, at least for a certain cases, more accurate than conventional goniometry, they similar to CAG cannot directly quantify local surface topography (e.g. area fraction) or surface chemistry (e.g. contact angle). 
However, previous studies in the literature using direct force measurement to characterise surfaces have failed to yield a quantitative prediction of area fraction from force measurements. For example, \citeauthor{paxson_self-similarity_2013} measured the detachment force between a droplet and a superhydrophobic surface with microscopic square pillars, and observed a monotonic increase in the detachment force with $\phi$ \cite{paxson_self-similarity_2013}. In contrast, \citeauthor{jiang_topography-dependent_2020} measured the maximum adhesion force between a droplet and superhydrophobic surfaces with cylindrical pillars and observed a non-monotonous increase in the maximum adhesion force with $\phi$  \cite{jiang_topography-dependent_2020}. 
%These seemingly opposite experimental observations make it important to study this phenomenon in more detail.
In this Letter, using numerical simulations, we demonstrate that microscale interfacial deformation near the contact line governs the force acting on a droplet during retraction from the surface. We present an experimentally validated expression relating adhesion force to pillar area fraction that can be used to quantitatively characterize a surface from just the value of the maximum adhesion force exerted by a droplet in the Cassie-Baxter wetting regime.

The net vertical force on a droplet in contact with a surface is $F = F_{\rm{p}}-F_{\sigma}$ where $F_{\rm{p}}$ is the pressure force due to the Laplace pressure difference ($\Delta p$) and $F_{\rm{\sigma}}$ is the surface tension (or pinning) force acting at the contact line (CL).
$F_{\rm{p}}$ acts downward, pushing the droplet against the surface, while $F_{\rm{\sigma}}$ exerts a pulling force, acting to detach the droplet. 
%$F_{\rm{\sigma}}$ is governed by the ability of the surface to effectively pin the CL. 
A surface with a higher pinning strength can pin the CL for a longer duration of time, during which the interface near the CL continuously deforms as the droplet is gradually pulled away from the surface. The relationship between force and interfacial deformation is crucial to developing a universal model for the adhesion force as a function of the surface topography. Microscopic imaging of moving CL is difficult, and it is not possible to capture multiple CL jump events that occur simultaneously at different points around the circumference of the droplet \cite{schellenberger_how_2016, paxson_self-similarity_2013, yeong_microscopic_2015, cheng_microscopic_2005, papadopoulos_how_2013, nosonovsky_patterned_2008, li_receding_2016, vieira2024through}. 
%This lack of understanding of the microscale dynamics of the interface hampers our ability to truly understand this relationship. 

To better understand the relationship between adhesion forces and surface topography, we employed the numerical technique presented in the accompanying paper \cite{kumar2025method} to simulate the microscale interfacial dynamics during a typical adhesion force measurement experiment.
%%%%%%%%%%%%%%%%%%%%%%%%%%%%%%%%%%%%%%%%%%%%%
%____Big Figure_____
\begin{figure}
    \centering
    \includegraphics[width=\linewidth]{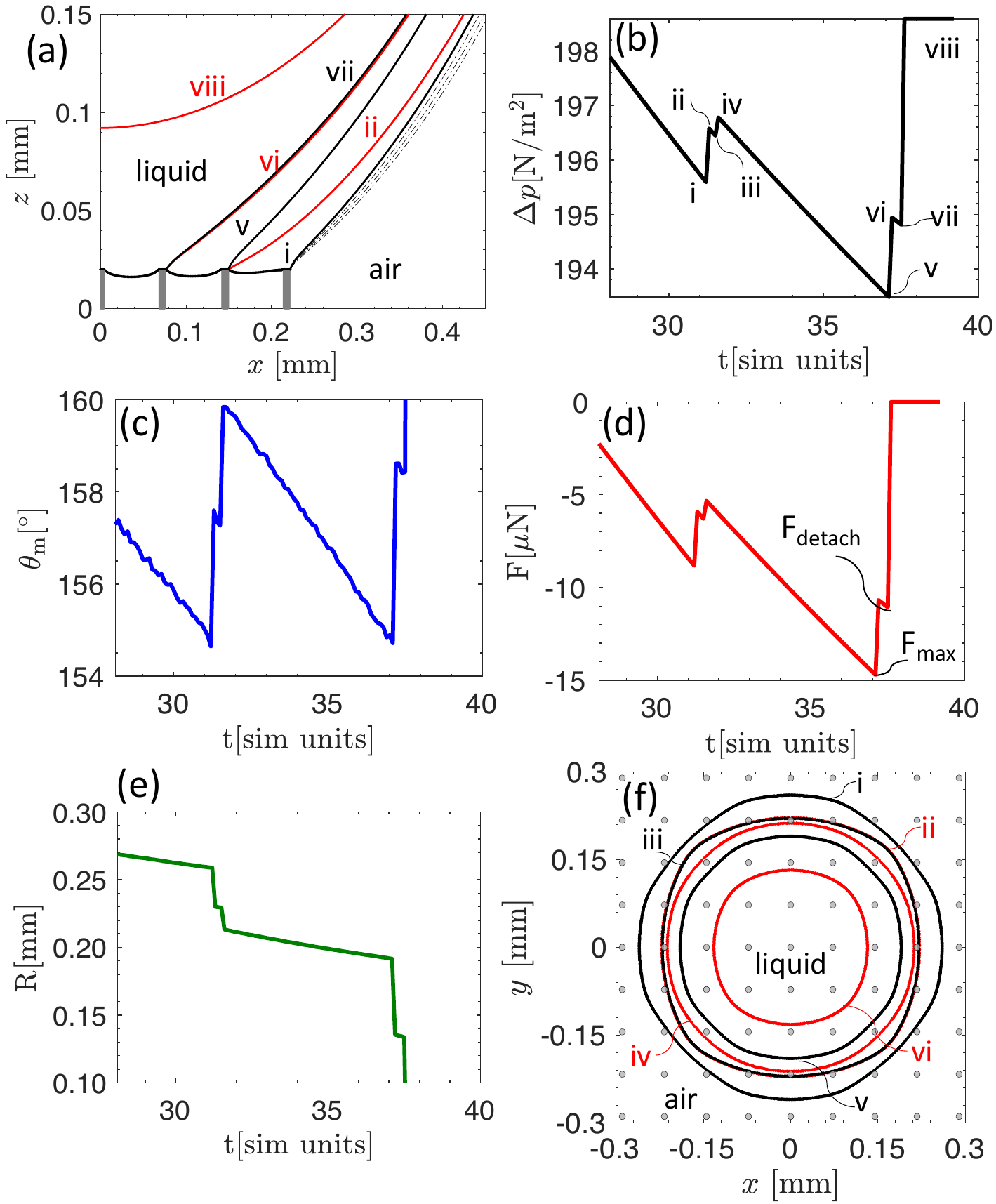}
    \caption{%(a) Equilibrium droplet morphology showing the meridional ($\kappa_1$) and azimuthal ($\kappa_2$) curvatures. 
    %$\kappa_1$ and $\kappa_2$ are a function of the macroscopic droplet radii in the principal planes.
    Simulation of the \textit{recede} stage for a droplet ($V = 1.6$ $\mu$L, $\theta_{\rm{A}} = 111.2$\textdegree, $\theta_{\rm{R}} = 98.7)$\tc on a superhydrophobic surface with $\phi = 0.015$, showing the (a) corresponding liquid-air interface equilibrium states and variation of (b) Laplace pressure ($\Delta p$), (c) contact angle ($\theta_{\rm{m}}$), (d) net vertical force $F$, (e) droplet base radius ($R$), and (f) apparent CL. Points $i,iii,v,vii$ and $ii, iv, vi$ and $viii$ represent  the equilibrium states just before and after the CL executes a jump, respectively. Solid black and red lines in (a) and (f) represent the first and the second critical states respectively. The dashed black lines in (a) represent the droplet profile when the CL is pinned on a set of pillars before executing jump $i \rightarrow ii$. Points $iii$ and $iv$ are not shown in (a) as the jump $iii \rightarrow iv$ occurs out of the plane. The droplet base radius and macroscopic contact angle are calculated at the apparent CL shown in (f) at 20 $\mu$m above the pillar tops.} 
    \label{fig:typical_variations}
\end{figure}
%____Big Figure ends_____
%%%%%%%%%%%%%%%%%%%%%%%%%%%%%%%%%%%%%%%%%%%%%
Fig. \ref{fig:typical_variations} shows the typical variation in droplet parameters as it is pulled away from the surface during the \textit{recede} stage \cite{kumar2025method}. Due to the CL pinning,
% The net force in the vertical direction ($F$) is the summation of the pressure force ($F_{\rm{p}}$) that pushes the droplet against the surface and the surface tension force ($F_{\sigma}$) pulls the droplet away from the surface, that is,
% %
% \begin{equation}
%     F = F_p - F_\sigma  
%     \label{eqn:force_equation}
% \end{equation}
% %
%When the droplet is pulled, that is, the \textit{recede} stage, due to the pining of the CL, 
the meridional curvature ($\kappa_1$, Fig. 1a in \cite{kumar2025method}) decreases (Fig. \ref{fig:typical_variations}a), and the Laplace pressure inside the droplet ($\Delta p)$ gradually decreases as the droplet is stretched (Fig. \ref{fig:typical_variations}b). The CL remains pinned until $\kappa_1$ reduces to a value at which the interface cannot exist in equilibrium, (ie. a constant mean curvature with Young's angle boundary condition at the CL),  %This decreases $\kappa_1$ until the interface cannot be in equilibrium - constant mean curvature with the Young's angle boundary condition at the CL - and%
at which the CL executes a jump (Fig. \ref{fig:typical_variations}a). The CL will pin again at a new location if the conditions for the droplet equilibrium are satisfied. 
%The equilibrium droplet morphologies just before and after a contact line jump event are, respectively, the first and second critical states \cite{kumar_energy_2024-1}. 
%In Fig. \ref{fig:typical_variations}c, the portion of the interface lying on the $x-z$ plane for some of the equilibrium states is shown.
%The first and second critical states corresponding to the points $p_1,q_1,r_1$ and $p_2,q_2,r_2$ in Fig. \ref{fig:typical_variations}b are shown as solid black and red lines. 
%The dashed lines in Fig. \ref{fig:typical_variations}c show the equilibrium droplet morphologies before the first critical state is reached and the CL executes the jump. 
During the CL jump, $\kappa_1$ increases slightly, 
depending on the distance the CL moves during the jump. This results in a slight increase in $\Delta p$, explaining the sawtooth variation in the Laplace pressure. During this stage, when the CL is pinned, the kinematics of the interface dictates a gradual decrease in $\theta_{\rm{m}}$ (measured at the apparent CL at a height $\Delta z$ above the pillars, see Fig. \ref{fig:drop_footprint}a). A decreasing $\Delta p$ and $\theta_{\rm{m}}$ results in an increasing magnitude of the force $F$, again leading to a sawtooth variation observed experimentally by \citeauthor{liimatainen_mapping_2017} \cite{liimatainen_mapping_2017}. The force reaches a maximum ($F_{\rm{max}}$) just before the droplet detaches from the surface at a smaller force $F_{\rm{detach}}$ (Fig. \ref{fig:typical_variations}d, see end matter \ref{SI:why_fmax}). The droplet base radius ($R$), measured as the radius of the apparent CL at $\Delta z = 20$ $\mu$m, moves in a step-like manner due to the CL pinning followed by occasional jumps (Fig. \ref{fig:typical_variations}e). 
The change in force due to a jump is higher at locations corresponding to $i$ and $iii$ in comparison to $v$, because these jumps occur in the direction of the surface structure, ie. in the $x$ (or $y$) axis direction. When a jump is aligned with surface structure the interface dissipates a greater magnitude of energy compared to when it jumps in the direction 45\tc to the surface structure \cite{kumar_energy_2024-1}, eg. for $v \rightarrow vi$ (Fig. \ref{fig:typical_variations}f).

A force balance on the drop gives an expression for the maximum force:
\begin{equation}
    F_{\rm{max}} = \Delta p A - \sigma\sum_{i=1}^n l_i \sin\theta_i 
    %\approx  \Delta p \pi R^2 - 2\pi R\sigma \sin \theta_{\rm{m}}
     \label{eqn:net_vertical_force}
\end{equation}
where $\sigma$ is the surface tension of the liquid-air interface, $l_i$ and $\theta_{\rm{i}}$ are the length and contact angle on the $i^{\rm{th}}$ pillar and $n$ is the total number of pillars under the droplet. 
 
%For a given droplet volume, both $F_{\rm{p}}$ and $F_{{\sigma}}$ depend on the footprint of the droplet. 
Eq. (\ref{eqn:net_vertical_force}) has two main challenges. Firstly, it is not possible to experimentally measure the length of the pinned contact line and the local contact angle on every pillar along the CL. Secondly, since the contact angle also varies around individual pillars \cite{li_receding_2016}, a suitable averaged contact angle is needed. A common assumption made in the literature is that the footprint is circular with a base radius $R$ \cite{jiang_droplet_2021,jiang_topography-dependent_2020,jiang_fluid_2024,sun_direct_2017,wang_contact_2020, daniel2023droplet}, simplifying Eq. (\ref{eqn:net_vertical_force}) to: 
\begin{equation}
    F_{\rm{max}}\approx  \Delta p \pi R^2 - 2\pi R\delta \sigma \sin \theta_{\rm{m}} 
\label{eqn:net_vertical_force2}
\end{equation}
where $\theta_{\rm{m}}$ is an average contact angle along the apparent CL and $\delta$ is the ratio between the length of the effective contact line and the apparent perimeter \cite{jiang_generalized_2019, xu_sticky_2012,jiang_droplet_2021,wang_contact_2020,sarshar_depinning_2019}. 
%However, Eq.(\ref{eqn:F_sigma_2}) will have inaccuracies if $\theta_{\rm{m}}$ and $R$ are not measured in the actual CL, where $\delta$ is defined. 
However, the footprint of the droplets is not circular. Fig. \ref{fig:drop_footprint}a shows a quarter of the droplet in equilibrium depicting the highly distorted liquid-air interface near the tops of the pillars. The actual CL is discrete, lying on the tops of the individual pillars. The interface under the droplet is almost flat, and therefore the only contribution to the force $F$ is provided by the pillars on the periphery of the droplet (Fig. \ref{fig:drop_footprint}a). 
%The concept of an effective contact line is often used to estimate $l_i$ from $R$ as $l_i = 2\pi R \delta$, where $\delta$ is the ratio between the length of the effective contact line and the apparent perimeter \cite{jiang_generalized_2019, xu_sticky_2012,jiang_droplet_2021}. 

In typical contact angle goniometry it is not possible to capture the microscopic contact angle and interfacial distortions near the actual CL, and instead the contact angle and base radius are measured at an apparent CL at a distance $\Delta z$ above the actual contact line at the pillars \cite{marmur_soft_2006, schellenberger_how_2016}, thus representing macroscopic quantities. The apparent CL varies according to $\Delta z$ as shown in Fig. \ref{fig:drop_footprint}b, and thus both $R$ and $\theta$ are functions of $\Delta z$ (Figs. \ref{fig:drop_footprint}c and \ref{fig:drop_footprint}b). Therefore, the droplet base area calculated from the experimentally measured droplet base radius differs from the actual droplet footprint, as shown in Fig. \ref{fig:drop_footprint}d. In addition, both $R$ and $\theta_{\rm{m}}$ depend on the viewing direction due to the distortion of the interface near the contact line (Fig \ref{fig:drop_footprint}b). %Here, $\psi$ is the viewing angle relative to the $x$ axis in the counter-clockwise sense. 
\begin{figure}
    \centering
    \includegraphics[width=\linewidth]{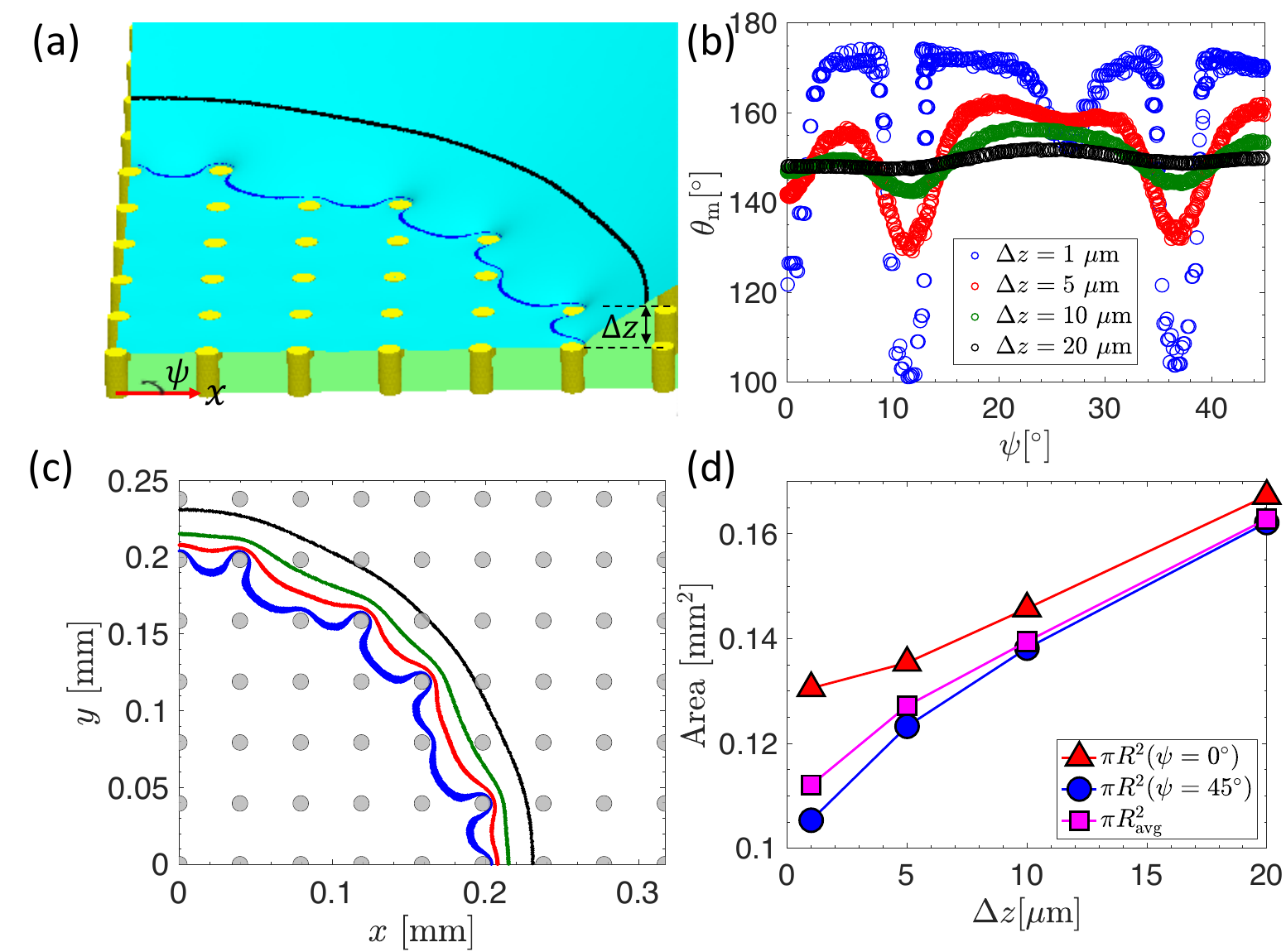}
    \caption{(a) Droplet morphology depicting one of the equilibrium states during a typical \textit{recede} stage for volume $V = 1.6$ $\mu$L, $\theta_{\rm{A}} = 111.2$\textdegree, $\theta_{\rm{R}} = 98.7$\tc on a superhydrophobic surface with $\phi = 0.05$. The CL at $\Delta z=1$ $\mu$m and 20 $\mu$m from pillar tops, respectively are shown in blue and black.
    %actual contact line and an apparent contact line, at $\Delta z$ distance from the pillar tops are shown in red and brown respectively. 
    (b) Macroscopic contact angle variation ($\theta_{\rm{m}}$) measured at different $\Delta z$ relative to the viewing direction ($\psi$) with respect to the $x$ axis. (c) Apparent CL at $\Delta z=1,5,10$ and 20 $\mu$m are shown in blue, red, green and black respectively. (d) Droplet footprint area as a function of $\Delta z$. The area is estimated considering a circular CL based on the base radius measured along the $x$ axis ($\psi=0$\textdegree) and at 45\tc to the $x$ axis are shown as filled triangles and circles respectively. The variation in the area based on the averaged base radius is also shown as filled squares.
    %Droplet footprint area as a function of $\Delta z$, based on an average value of the base radius %normalised by the apparent footprint area calculated numerically using the CL shape at $\Delta z=1$ $\mu$m. 
    %Estimated footprint area considering a circular CL is also shown based on the base radius measured along the $x$ axis ($\psi=0$\textdegree) and at 45\tc to the $x$ axis. 
    The results shown are for $\phi=0.05$.}
    \label{fig:drop_footprint}
\end{figure}
%__________________%
%where $\theta_{\rm{m}}$ is an average contact angle along the apparent CL. However, Eq.(\ref{eqn:F_sigma_2}) will have inaccuracies if $\theta_{\rm{m}}$ and $R$ are not measured in the actual CL, where $\delta$ is defined. 
%As an approximation, we can assume that the $R$ and $\theta_{\rm{m}}$ measured at an apparent CL are the values at the actual CL and can estimate the net vertical force with $\delta=1$:
%____________Eq.4________________%
%
% \begin{equation}
%     F = \Delta p \pi R^2 - 2\pi R\sigma \sin \theta_{\rm{m}}
%     \label{eqn:net_vertical_force2}
% \end{equation}
%
%____________Eq.4________________%
%Eq. (\ref{eqn:net_vertical_force}) is shown in Fig. \ref{fig:drop_footprint}e together with the numerically predicted results for a surface with $\phi=0.05$.  

Despite the limitations discussed above, Eq. (\ref{eqn:net_vertical_force2}) can be used to estimate the force experimentally by measuring $R$ and $\theta_{\rm{m}}$ using CAG, with $\delta$ as a fitting parameter. However, to our knowledge, there is no direct method for estimating $\delta$ \cite{jiang_topography-dependent_2020, jiang2019generalized}. We observe that Eq. (\ref{eqn:net_vertical_force2}) with $\delta=1$, slightly underestimates the forces (see \S \ref{SI:Eq2} in the end matter). In addition, Eq. (\ref{eqn:net_vertical_force2}) predicts a nearly similar force variation irrespective of $\Delta z$ as long as both $R$ and $\theta_{\rm{m}}$ are measured at the same $\Delta z$ (refer to \S \ref{SI:Eq2} in the end matter). 
However, Eq. (\ref{eqn:net_vertical_force2}) cannot be used to estimate the area fraction of the surface from force data.
%conventional goniometry is unable to capture the microscopic contact angle and the interfacial distortions near the actual CL. 

\begin{figure}
    \centering
    \includegraphics[width=\linewidth]{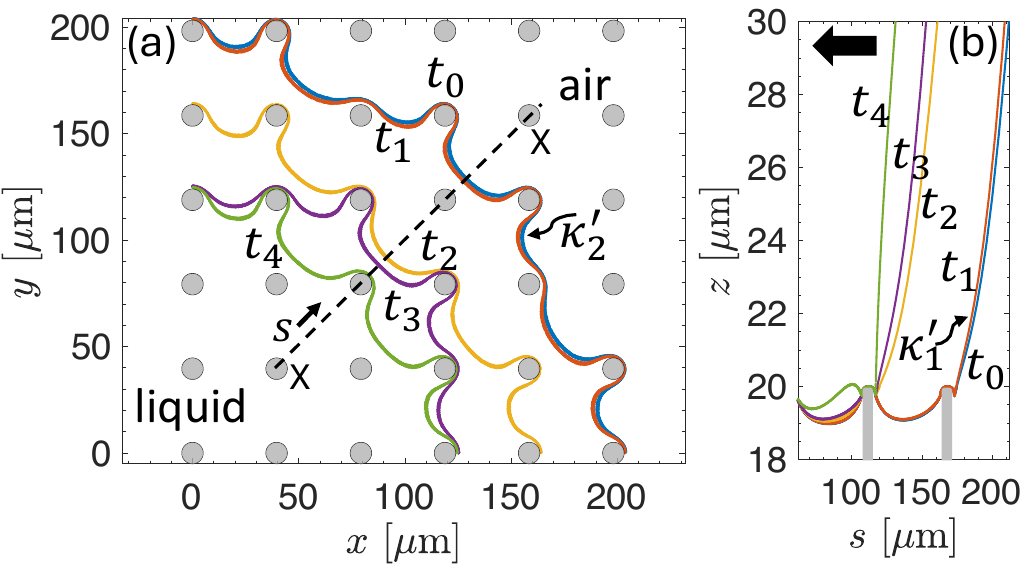}
    \caption{Apparent CL ($\Delta z=1$ $\mu$m) at different times during \textit{recede} stage for a droplet of volume $V = 1.6$ $\mu$L, $\theta_{\rm{A}} = 111.2$\textdegree, $\theta_{\rm{R}} = 98.7$\tc on a superhydrophobic surface with $\phi = 0.05$ ($t_4>t_3>t_2>t_1>t_0$). (b) Liquid-air interface in the plane corresponding to the $X-X$ section in (a). %depicting the CL pinning and the first CL jumping event.
    }
    \label{fig:microscale}
\end{figure}

%We now discuss the methodology to relate the results of the direct force measurement to the surface topography without using any other parameters such as the contact angle and the droplet base radius. 
To understand the effect of roughness density on the net vertical force, we need to look at the microscale dynamics of the interface. Near the CL, due to separation of length scales, $l_{\rm{c}}\ll R$ (where $l_{\rm{c}}$ is the roughness length scale), the pressure difference across the interface is negligible, and the interface behaves as a minimal surface (zero mean curvature). As the droplet is stretched macroscopically, the elastic energy is stored in microscale deformations \cite{joanny_model_1984, kumar_energy_2024-1}. Due to these deformations, both the meridional and azimuthal curvatures of the interface change locally to maintain a zero mean curvature at every point of the interface in the microscopic region. Fig. \ref{fig:microscale}a shows the apparent CL ($\Delta z=1$ $\mu$m) at different times during the \textit{recede} stage on a superhydrophobic surface with $\phi=0.05$. At $t_0$, the CL is pinned while the droplet is pulled macroscopically. The local curvature in the azimuthal plane ($\kappa_2^{'}$) increases to maintain the condition of a zero mean curvature. At $t_1$, at certain locations, $\kappa_2^{'}$ reaches the maximum value that can be supported by the pillars, and the CL executes a jump. %It is worth noting that the local curvature varies around the CL. 
Pillars that are closer can support a higher value of $\kappa_2^{'}$ compared to those that are sparse \cite{kumar_energy_2024-1}. The CL jump begins locally where the condition of zero mean curvature cannot be satisfied depending on the local surface topography. The CL executes a jump (at $t_1$) and finds a new configuration in the subsequent rows of the pillars. As the droplet recedes further, the CL jumps at $t_2$ and $t_3$ and detaches from the surface at $t_4$. 
%In Fig. xxx b we show the apparent CL of some of the equilibrium states of the droplet during the \textit{recede} stage. We can observe the increase in local azimuthal curvature ($\kappa^{'}_2$) as the droplet is stretched. 
In Fig. \ref{fig:microscale}b, a corresponding increase in the local meridional curvature ($\kappa^{'}_1$) can be observed from the cross-section of the interface. The increase in $\kappa_1^{'}$ corresponding to $t_0$ and $t_1$ can be seen. For a given surface chemistry (Young's angle), the magnitude of the azimuthal curvature depends on the local surface topography, with closely spaced heterogeneities resulting in a curvature higher than that of sparsely spaced. Therefore, superhydrophobic surfaces with a high pillar area fraction will provide a greater pinning force and, therefore, higher maximum and detachment forces. 

\begin{figure}
    \centering
    \includegraphics[width=0.80\linewidth]{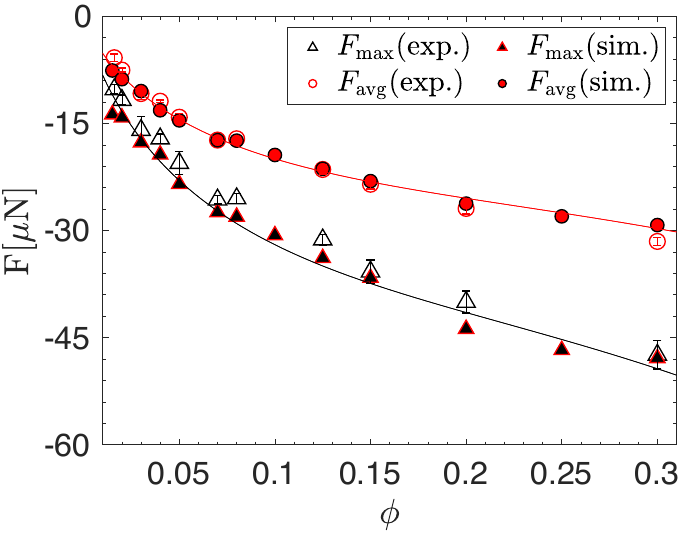}
    \caption{Variation in the maximum adhesion force ($F_{\rm{max}}$) and average tensile force ($F_{\rm{avg}}$) with pillar area fraction ($\phi$) based on numerical simulations and physical experiments, for a droplet volume $V = 1.6$ $\mu$L, $\theta_{\rm{A}} = 111.2$\textdegree, $\theta_{\rm{R}} = 98.7$\tc and $d=10$ $\mu$m,  are shown by filled and empty symbols respectively. Eq. (\ref{eqn:non_dilute_fit}) using the numerically predicted values of the parameters $a,b,c$ are also shown as black and red curves for $F_{\rm{max}}$ ($a=331.8,b=-756.3,c=400.5$) and $F_{\rm{avg}}$ ($a=209.1,b=-458.7,c=253.2$) respectively.}
    \label{fig:force_vs_phi}
\end{figure}

The adhesion force is proportional to the elastic energy originating from interfacial deformations. Using the arguments of \citeauthor{joanny_model_1984} \cite{joanny_model_1984} for a dilute surface, as well as incorporating interactions between neighbouring pillars at higher area fractions \cite{kumar_energy_2024-1}, the adhesion force can be written as,
\begin{equation}
    F/\sigma d = a\phi\ln\phi + b\phi^2 + c\phi.  
    \label{eqn:non_dilute_fit}
\end{equation}
where $d$ is the pillar diameter; and $a$, $b$ and $c$ are, for a fixed droplet volume, constants depending on the geometry of the pillars and the inherent receding contact angle on a chemically similar but flat surface ($\theta_{\rm{R}}$). In Fig. \ref{fig:force_vs_phi} we show the variation in $F_{\rm{max}}$ with $\phi$ obtained by energy minimization simulations on superhydrophobic pillared surfaces (10 $\mu$m in diameter and 20 $\mu$m high) with inherent advancing and receding contact angles $\theta_{\rm{A}}=111.2$\tc and $\theta_{\rm{R}}=98.7$\tc respectively. 
% A fitting equation of the form given in Eq. (\ref{eqn:non_dilute_fit}) with constants $a=331.842,b=-756.334,c=400.464$
% %$a=241.581,b=-550.611,c=291.538$, this is dimensional parameters
% is also shown as a solid black curve. 
Good agreement is observed between the fit of Eq. \ref{eqn:non_dilute_fit} using $F_{\rm{max}}$ and the simulation data, especially at low area fractions. However, at higher area fractions, the fit starts deviating from the numerically predicted values, potentially due to the discrete nature of $F_{\rm{max}}$ on a rough surface. As the CL recedes on the surface, $F_{\rm{max}}$ represents just one of the many equilibrium CL configurations throughout the receding motion and may not be suitably representative of the average surface properties. 

Instead, we present an alternative metric; the average tensile force $F_{\rm{avg}}$, which is the average of the force on the droplet for $F<0$ (according to the sign convention used in this study). Unlike $F_{\rm{max}}$, $F_{\rm{avg}}$ depends on an average value of $\theta_{\rm{m}}$ and $R$ which is more representative of the receding motion of the CL. The fit to $F_{\rm{avg}}$ simulation data is shown in Fig. \ref{fig:force_vs_phi}, showing a better agreement over the full range of $\phi$ considered.

To confirm this experimentally, we measured the adhesion force on superhydrophobic surfaces with cylindrical pillars of diameter 10 $\mu$m and height 20 $\mu$m (etched onto a silicon wafer using a dry etching technique \cite{kumar_energy_2024}). Before force measurement, thin layers of chromium (5 nm) and gold (25 nm) were deposited by physical vapour deposition, and then the surfaces were hydrophobized with a monolayer of octadecanethiol from a 1 mM ethanol solution (\cite{kumar_energy_2024}).
%Before force measurement, the surfaces were hydrophobized by depositing a thin monolayer of octadecanethiol (\cite{kumar_energy_2024}). 
The advancing and receding contact angles on flat hydrophobized surfaces were measured to be 111.2\tc and 98.7\textdegree, respectively (\cite{kumar_energy_2024}). A Dataphysics DCAT-21 tensiometer was used to measure the adhesion between the surface and a 1.6 $\mu$L droplet of ultra pure DI water (less than 4 ppm total organic carbon and an interfacial tension of 72.8 mN/m at 20 $^{\circ}$C, see Appendix \ref{SI:exp}). There is excellent agreement between the simulation and experiment. It is also clear that the fit of Eq. \ref{eqn:non_dilute_fit} to $F_{\rm{avg}}$ is better than the fit to $F_{\rm{max}}$, and is therefore better representative of the surface topography. 
%In addition, Eq. (\ref{eqn:non_dilute_fit})  agrees well with the experimentally measured $F_{\rm{max}}$ and $F_{\rm{avg}}$, with a better correlation observed for the values $F_{\rm{avg}}$. 

An important point to note here is that, in all the experiments conducted in this study, droplet evaporation was negligible due to the relatively high droplet receding velocity (the stage upon which the surface is placed was moved away from the droplet at a speed 25 $\mu$m/s). 
%With an uncontrolled droplet evaporation, the vapours can collect in the crevices between the pillars. These vapours will exert an additional pressure ($p_{\rm{vap}}$) opposite to the Laplace pressure inside the droplet (for convex droplets) as shown in Fig. \ref{fig:evaporation}a. This reduces the pressure force contribution pushing the droplet against the surface, resulting in an early droplet depinning and hence a lower magnitude of $F_{\rm{max}}$. To validate this hypothesis, 
To determine the role of evaporation in the measurement, we conducted experiments measuring the adhesion force between a droplet 1.6 $\mu$L and pillared surfaces with area fractions ranging from 0.016 to 0.30. The droplet was allowed to evaporate and the stage was kept stationary. The $F_{\rm{max}}$ values are shown in Fig. \ref{fig:SI_evap} as open squares. We observed a general increase in $F_{\rm{max}}$ with $\phi$, but without any universal relationship between them. Similar observations were made by \citeauthor{jiang_topography-dependent_2020} \cite{jiang_topography-dependent_2020}. 
%We suspect that this random variation in $F_{\rm{max}}$ is due to the randomness of the vapour accumulation underneath the droplet. 
We also conducted numerical simulations that mimic droplet evaporation %but neglect vapour accumulation 
(see \S IV C in \cite{kumar2025method}). The simulation results are shown in Fig. \ref{fig:SI_evap} as triangles and show a monotonous increase with $\phi$ that can be predicted by Eq. (\ref{eqn:non_dilute_fit}) which is shown as a solid red curve. The discrepancy between the simulation results and the experimental data is due to the change in droplet weight as a result of evaporation. The tensiometer apparatus is usually calibrated with the initial weight of the droplet, and any change in it due to evaporation results in under-prediction of $F_{\rm{max}}$ (see \S IV D in \cite{kumar2025method}). To incorporate the difference in droplet weight due to evaporation, we plot $F_{\rm{max}}+\Delta W$ with $\phi$ in Fig. \ref{fig:SI_evap} as circles. Here, $\Delta W$ is the change in droplet weight during the \textit{recede} stage. The maximum force considering the change in droplet weight due to evaporation during \textit{recede} shows a monotonous increase with $\phi$ that can be represented by Eq. (\ref{eqn:non_dilute_fit}) (green curve in Fig. \ref{fig:SI_evap}) and is in better agreement with the experimental data. However, the simulation assumes negligible evaporation of the droplet during the \textit{advance} stage, which may not be true for the experiments, and explains the discrepancy between the two.
%%possibly due to the presence of vapors near the CL. 
%The simulation only considers the effect of a change in droplet volume due to evaporation. This requires careful experiments with controlled evaporation rate, which is beyond the scope of this work. 
In addition, the numerical results form an upper bound to the experimental data, re-affirming the fact that droplet evaporation results in a reduction in $F_{\rm{max}}$. %Finally, to rule out the possibility that a slightly different initial droplet volume could have caused these random variations in $F_{\rm{max}}$, we used a slightly larger droplet (3 $\mu$L) and measured adhesion at a rate of 25 $\mu$m$/$s. The $F_{\rm{max}}$ for the larger droplets are shown as black circles in Fig. \ref{fig:SI_evap}a. We observe that the difference in $F_{\rm{max}}$ due to a change in the initial droplet volume (negligible evaporation) is much smaller than what is observed when the initial droplet volume remains unchanged, but the droplet evaporation is significant. 
Therefore, an important criterion for employing the direct force measurement method as a surface characterization tool is to minimize the effect of droplet evaporation. This also explains the different observations made by \citeauthor{paxson_self-similarity_2013} \cite{paxson_self-similarity_2013} where the evaporation of the droplets was controlled and \citeauthor{jiang_topography-dependent_2020} \cite{jiang_topography-dependent_2020} where the evaporation was not controlled.

To conclude, we have shown that the average pulling force exerted by a droplet for a given droplet volume, pillar geometry, and inherent receding angle during typical droplet probe microscopy when evaporation is negligible, is a function of pillar area fraction only and  can be represented by Eq. (\ref{eqn:non_dilute_fit}). The coefficients in the fitting equation can be fully predicted using the numerical technique presented in the accompanying paper \cite{kumar2025method}. The experimentally measured average force along with the numerically predicted coefficients in Eq. (\ref{eqn:non_dilute_fit}) are sufficient to quantitatively characterize a superhydrophobic pillared surface.

\begin{acknowledgments}
J.D.B. received support from an Australian Research Council Future Fellowship (FT220100319) funded by the Australian Government. This work was performed in part at the Materials Characterisation and Fabrication Platform (MCFP) at The University of Melbourne.
\end{acknowledgments}

\bibliography{ref_drop_adhesion.bib}% Produces the bibliography via BibTeX.

\clearpage
\appendix
\section{Variation in pressure and surface tension forces during the \textit{recede} stage}
\label{SI:why_fmax}

\begin{figure}[h]
    \centering
    \includegraphics[width=0.80\linewidth]{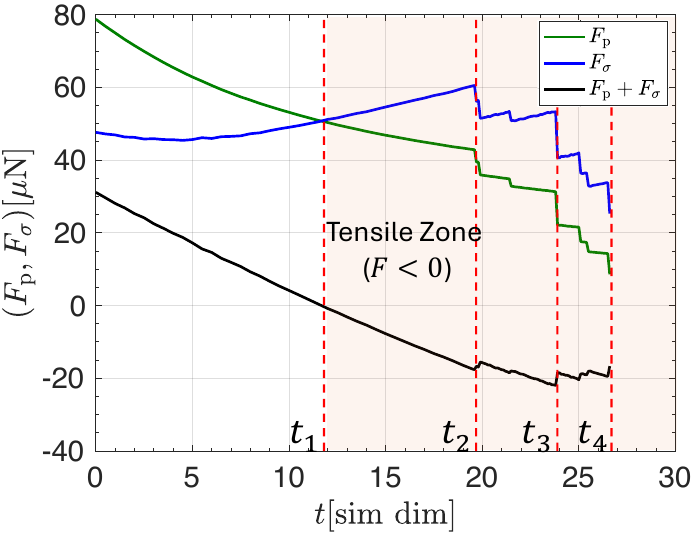}
    \caption{Variation in the pressure ($F_{\rm{p}}$), surface tension ($F_{\sigma}$) and total force ($F=F_{\rm{p}}+F_{\sigma}$) with time (t) in simulation units. The simulations were performed for a droplet volume $V=1.6$ $\mu$L, $\theta_{\rm{A}}=111.2$\tc and $\theta_{\rm{R}}=98.7$\tc on a superhydrophobic surface with $\phi=0.05$.}
    \label{fig:SI_fig1}
\end{figure}

Fig. \ref{fig:SI_fig1} shows the variation in pressure ($F_{\rm{p}}$), surface tension ($F_{\sigma}$) and total force ($F_{\rm{p}}+F_{\sigma}$) with time (in simulation units). Initially, $F_{\rm{p}}$ is relatively higher than $F_{\sigma}$. Up to $t=t_2$, the CL is pinned and $F_{\sigma}$ and $F_{\rm{p}}$ show an increasing and decreasing trend, respectively, due to a decreasing trend $\theta_{\rm{m}}$ and $\Delta p$. At $t=t_1$, $F_{\sigma}$ and $F_{\rm{p}}$ are equal in magnitude but opposite in direction, and therefore the net force ($F=F_{\rm{p}}+F_{\sigma}$) is zero. Further stretching of the droplet results in a negative net vertical force or a tensile (pulling) force on the surface. The tensile region is shown as the shaded area in the figure. The CL starts to recede at $t=t_2$, executing a jump which results in a drop in both $F_{\rm{p}}$ and $F_{\sigma}$. Now, since $F_{\rm{p}}$ varies with the square of the base radius, it drops faster than $F_{\sigma}$, which depends linearly on the base radius. This gives the force curve its peculiar form in which the maximum force ($F=F_{\rm{max}}$ at $t=t_3$) is achieved before the droplet detaches from the surface at a comparatively smaller force ($F=F_{\rm{detach}}$ at $t=t_4$).

\section{Effect of the distance at which macroscopic droplet parameters are measured}
\label{SI:Eq2}

%%%%%%%%FIG2%%%%%%%%%%
\begin{figure}[h]
    \centering
    \includegraphics[width=0.80\linewidth]{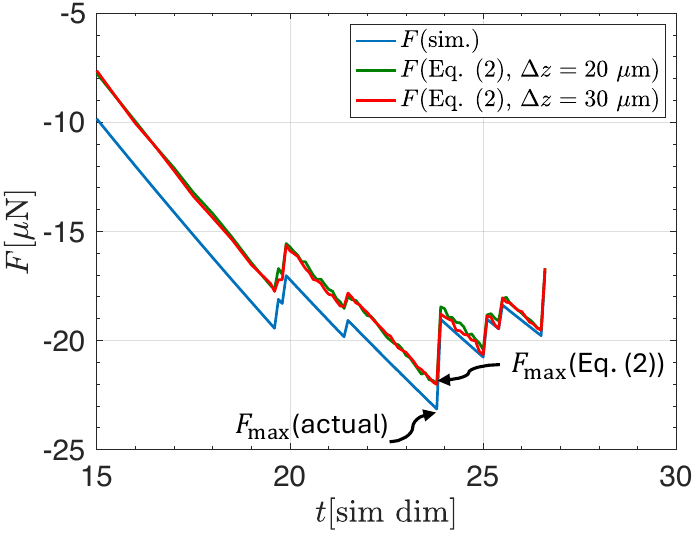}
    \caption{Variation in the net vertical force ($F$) with time ($t$), in simulation units. The actual force based on simulations, and approximate force based on Eq. (\ref{eqn:net_vertical_force2}) in the main text are shown. The approximate force is calculated using the macroscopic droplet parameters obtained at $\Delta z=20$ $\mu$m and 30 $\mu$m respectively. Eq. (\ref{eqn:net_vertical_force2}) under-predicts the $F_{\rm{max}}$ by approximately 1.2 $\mu$N for a droplet volume $V=1.6$ $\mu$L, $\theta_{\rm{A}}=111.2$\tc and $\theta_{\rm{R}}=98.7$\tc on a superhydrophobic surface with $\phi=0.05$.}
    \label{fig:SI_fig2}
\end{figure}
Fig. \ref{fig:SI_fig2} shows the variation in the net vertical force ($F$) with simulation time during a typical \textit{recede} stage on a surface with $\phi=0.05$. The force variation obtained by using Eq. (\ref{eqn:net_vertical_force2}) in the main text slightly underestimates $F_{\rm{max}}$, which is $\sim 1.2$ $\mu$N for the case under consideration ($\phi=0.05$). However, Eq. (\ref{eqn:net_vertical_force2}) gives similar results for both $\Delta z=20$ $\mu$m and $\Delta z=30$ $\mu$m, respectively. Therefore, Eq. (\ref{eqn:net_vertical_force2}) can be used to predict the droplet adhesion forces from the macroscopically measured contact angle and the droplet base radius without using the concept of effective contact line length, as long as both $\theta_{\rm{m}}$ and $\Delta z$ are measured at the same $\Delta z$.

\section{Direct force Measurement}
\label{SI:exp}

\begin{figure}
    \centering
    \includegraphics[width=0.80\linewidth]{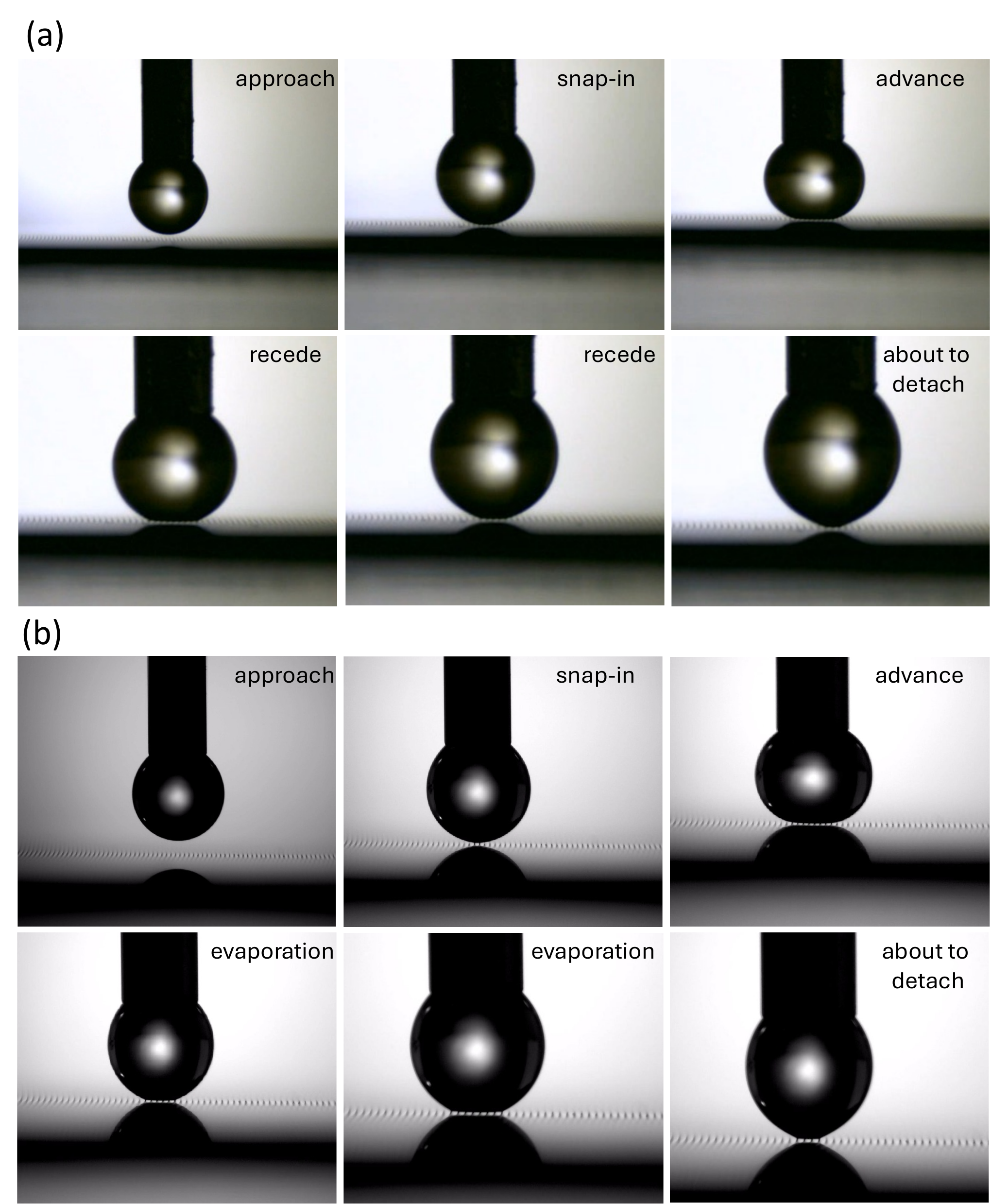}
    \caption{Snapshots of the direct force measurement between a 1.6 $\mu$L droplet ($\theta_{\rm{A}}=111.2$\tc and $\theta_{\rm{R}}=98.7$\textdegree) and superhydrophobic surface ($\phi=0.016$) when (a) the stage is moved downwards (25 $\mu$m/s) to initiate the CL receding, (b) the stage is kept stationary and the droplet evaporation causes the CL receding.}
    \label{fig:SI_recede}
\end{figure}

Direct force measurement experiments were conducted using the DCAT-21 tensiometer (DataPhysics). First, a superhydrophobic pillared surface, which was a $1.5 \times 1.5$ cm$^2$ piece, is gently placed on top of a movable stage. A 0.91 mm outer diameter needle is fixed to clamp attached to the tensiometer, for force measurement. In the present study, a droplet of precise volume (1.6 $\mu$L) of DI water (Milli-Q) was generated using a syringe pump, 
%[confirm with Marta],
which was then attached to the tip of the needle. The needle tip was tightly packed with Blu Tack to avoid water entrainment. During the \textit{advance} stage, the stage is moved at a speed of 50 $\mu$m/s until the droplet contacts the surface and the force increases to approximately 32 $\mu$N. Once the maximum compression force is reached, the stage is moved downward at a speed of 25 $\mu$m/s until the droplet detaches from the surface. The tensiometer measures the force between the droplet and the surface at a frequency of 50 Hz. The droplet was imaged simultaneously using a DinoLite camera (Edge Plus AM8917MZTL) to record the contact angle of the droplet and the diameter of the base (see Fig. \ref{fig:SI_recede}a).

%%%%%%%%FIG4%%%%%%%%%%
% \begin{figure}
%     \centering
%     \includegraphics[width=\linewidth]{figures/evaporation-sequence.pdf}
%     \caption{Snapshots of the direct force measurement between a 1.6 $\mu$L droplet ($\theta_{\rm{A}}=111.2$\tc and $\theta_{\rm{R}}=98.7$\textdegree) and superhydrophobic surface ($\phi=0.016$) when the stage is kept stationary and the droplet evaporation causes the CL receding.}
%     \label{fig:SI_evap}
% \end{figure}
% %
A second set of experiments was also conducted to see the effect of droplet evaporation. In these experiments, the stage was moved upwards at a speed of 10 $\mu$m/s until a compression force of 32 $\mu$N was reached. After this, the stage was kept stationary and the droplet was allowed to evaporate. The force between the droplet and the surface during this was measured by the tensiometer at a frequency of 50 Hz. During this process, droplet video was also recorded to estimate the macroscopic contact angle and the diameter of the droplet base (see Fig. \ref{fig:SI_recede}b). 
%%%%%%%%FIG3%%%%%%%%%%
\section{Effect of droplet evaporation}
\label{SI:evap}
\begin{figure}[h]
    \centering
    \includegraphics[width=0.70\linewidth]{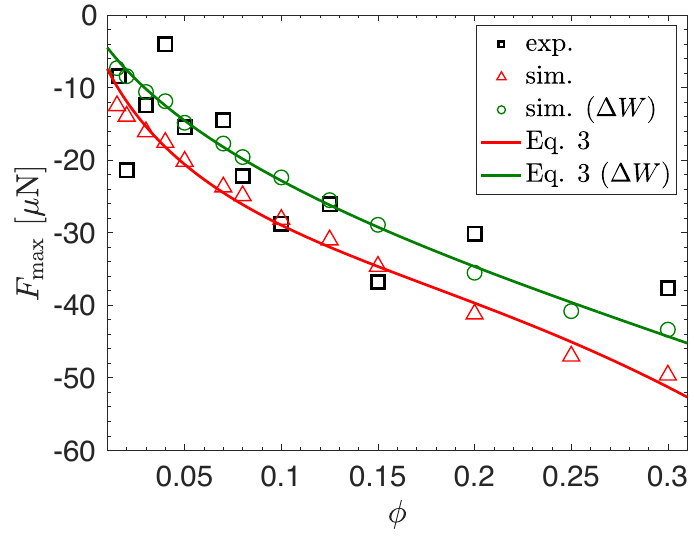}
    \caption{Variation in the maximum (red squares) %and average (black crosses) 
    force measured experimentally on an evaporating droplet of 1.6 $\mu$L volume with the pillar area fraction ($\phi$) on a superhydrophobic surface with $\theta_{\rm{A}}=111.2$\tc and $\theta_{\rm{R}}=98.7$\textdegree. The stage is kept stationary during the entire recede stage and droplet evaporation takes control of the CL depinning and droplet detachment. The simulation results ($F_{\rm{max}}$) of an evaporating droplet neglecting the change in droplet weight due to evaporation are shown as red triangles and the values incorporating the change in droplet weight ($\Delta W$) are shown as green circles. %Experimentally measured $F_{\rm{max}}$ for a 3.0 $\mu$L droplet is also shown as black circles. These results were obtained by moving the stage at 25 $\mu$m/s during the recede stage such that the effect of droplet evaporation is negligible. 
    Eq. (\ref{eqn:non_dilute_fit}) for both the cases, that is with and without considering change in droplet weight are also plotted as solid green and red curves respectively. %(b) Schematic of a droplet on a pillared superhydrophobic surface during a typical vertical force measurement experiment. The effect of droplet evaporation is shown. Vapors accumulates in the crevices on the surface resulting in a pressure $p_{\rm{vap}}$ acting opposite to $\Delta p$ causing an early onset of CL depinning.
    }
    \label{fig:SI_evap}
\end{figure}

Fig. \ref{fig:SI_evap} shows the variation in the maximum adhesion force with the pillar area fraction, measured experimentally for an evaporating droplet ($V=1.6$ $\mu$L, $\theta_{\rm{A}}=111.2$\textdegree, $\theta_{\rm{R}}=98.7$\textdegree). The simulation results neglecting any change in droplet weight and considering the change in droplet weight due to evaporation are also plotted as a scatter. The fittings based on Eq. (\ref{eqn:non_dilute_fit}) are also plotted as solid curves.
%and average adhesion force with the area fraction ($\phi$). Fig. \ref{fig:SI_evap}b shows the schematic of an evaporating droplet. With an uncontrolled droplet evaporation, the vapors can collect in the crevices between the pillars. These vapors will exert an additional pressure ($p_{\rm{vap}}$) opposite to the Laplace pressure inside the droplet (for convex droplets) as shown in the figure. This reduces the pressure force contribution pushing the droplet against the surface, resulting in an early droplet depinning and hence a lower magnitude of $F_{\rm{max}}$.
%

\end{document}